\documentclass{article}

\usepackage[preprint]{neurips_2026}

\usepackage[utf8]{inputenc}
\usepackage[T1]{fontenc}
\usepackage{hyperref}
\usepackage{url}
\usepackage{booktabs}
\usepackage{amsfonts}
\usepackage{nicefrac}
\usepackage{microtype}
\usepackage{xcolor}
\usepackage{amsmath}
\usepackage{amssymb}
\usepackage{graphicx}
\usepackage{bm}
\usepackage{bbm}
\usepackage{graphicx}
\usepackage{subcaption}
\newcommand{\ours}{GenTTP }
\newcommand{\ibours}{(GenTTP) }

\title{Generalising Travel Time Prediction To Varying Route Choices In Urban Networks}

\author{
  \textbf{Łukasz Gorczyca} \\
  Jagiellonian University \\
  Doctoral School of Exact and Natural Sciences \\
  \texttt{lukasz.gorczyca@uj.edu.pl}
  \And
  Kacper Drozd \\
  Jagiellonian University \\
  \texttt{kacper.drozd@uj.edu.pl} \\
  \AND
  Michał Bujak \\
  Jagiellonian University \\
  \texttt{michal.bujak@uj.edu.pl} \\
  \And
  Rafał Kucharski \\
  Jagiellonian University \\
  \texttt{rafal.kucharski@uj.edu.pl}
}

\begin{document}

\maketitle

\begin{abstract}
Previous methods that predict system-wide travel time, predominantly grounded in graph neural networks, remain limited to typical and recurring demand patterns. While they successfully predict future congestion following daily commute, they inherently approximate a single demand realisation and fail to capture varying route choices. In this work, we propose a Generalised Travel Time Predictor (GenTTP) that successfully differentiates route choices and offers accurate flow and travel time predictions. Our framework learns to uncover complex spatiotemporal traffic patterns and microscopic relationships between route choices and the resulting travel times. This addresses a critical gap: the lack of travel time prediction models that generalise across varying route assignments, where the same demand can produce substantially different network-wide outcomes depending on how travellers are distributed over available paths.
\end{abstract}

\section{Introduction}
System performance (total travel time) in urban traffic networks depends on two main factors. First, a path chosen by a driver (agent) that navigates her from the origin to the destination. Second, traffic that arises along the route. Naturally, when we look at the perspective of the whole urban network, the level of congestion at individual segments of the road depends on overlapping paths taken by agents. In such systems, to model the individual travel time, one can apply microscopic simulators (e.g. \texttt{Vissim}, \texttt{Aimsun} \cite{Vissim, aimsun}). Those methods simulate, second by second, the behaviour of individual agents and provide accurate estimates. However, due to the multi-agent microscopic nature, they are computationally heavy and stochastically unreliable for various large-scale applications. 

To mitigate these limitations, researchers \citep{Diff, WaveNet} proposed machine learning methods trained to provide quick estimates. Current methods \citep{When, STGNN, zheng2020gman} observe the state of the traffic network, infer drivers' likely movement patterns, and approximate their travel times. Such approach improves computation (replacing a 5-minute long simulation with a quick 0.01s inference); however, at the cost of precision and generalisability. They typically assume that agents follow a single and repetitive route choice pattern. Consequently, they fail when route choices differ (e.g. due to a disruption or an unexpected event). The changes in the traffic flow are then, typically flattened, and the model's goal is to provide a single estimate regardless of additional information on the system that may be available.

In particular, current methods fail to assess impact of collectively routing fleets (assigning routes to individual vehicles) of connected autonomous vehicles (CAVs), where the major task is not to learn the routing pattern, but to analyse the congestion under the specific pattern.
With the rising penetration rate of CAVs (fraction of all vehicles), differentiating routing strategies is an increasingly important task \citep{URB}, which, due to its complexity, remains unsolved. 
Fleet operators, who design routing patters (referred to as \textit{assignments}), will be able to control a significant part of traffic flow. 
To coordinate their vehicles, they need to recognise existing congestion, but also, crucially, to prevent it from emerging by their fleet itself. 
Our method addresses the key missing component in hitherto developed machine learning frameworks: capability to predict congestion and resulting travel time for various (often atypical) assignments.

To this end, we propose the Generalised Travel Time Predictor \ibours. 
Our method combines two main building branches that are fused with MLP into the final prediction.
The first branch, similarly to existing models, focuses on recognising the existing congestion. 
We aggregate flow information and, by applying spatio-temporal convolutions, obtain an aggregated representation.
We enrich our model with a classifier that recognises unoccupied road segments (existing methods tend to overestimate flows on them, e.g. \cite{WaveNet}).
The second branch of \ours is dedicated to analysing the assignment.
We process the information with GCN encoder and LSTM to obtain meaningful representation of the directional traffic propagation. 
Thanks to such architecture, our model successfully combines microscopic phenomena with microscopic trends.

\begin{figure}[htbp]
    \centering
    \caption{Overview of \ours framework. While existing graph-based predictors successfully capture recurring congestion patterns, they are observational and fail under distribution shifts, when routing policies vary. \ours predicts travel times conditioned on route patterns, enabling robust counterfactual reasoning. \ours is equipped with standard spatio-temporal GNN block for processing of traffic states and introduced assignment module designed for recognising impact of varying route choice patterns on future traffic states. These branches are followed by a fusion layer, which gives system-wide travel time as an output. In comparison to baseline traffic flow predictors, which are not able to process routing information, \ours can predict travel time assigned to path varying trips, while baselines give only amenable output from across patterns.}
    \includegraphics[scale=0.5, width=1\linewidth, trim = {0cm, 0cm, 0cm, 0}, clip]{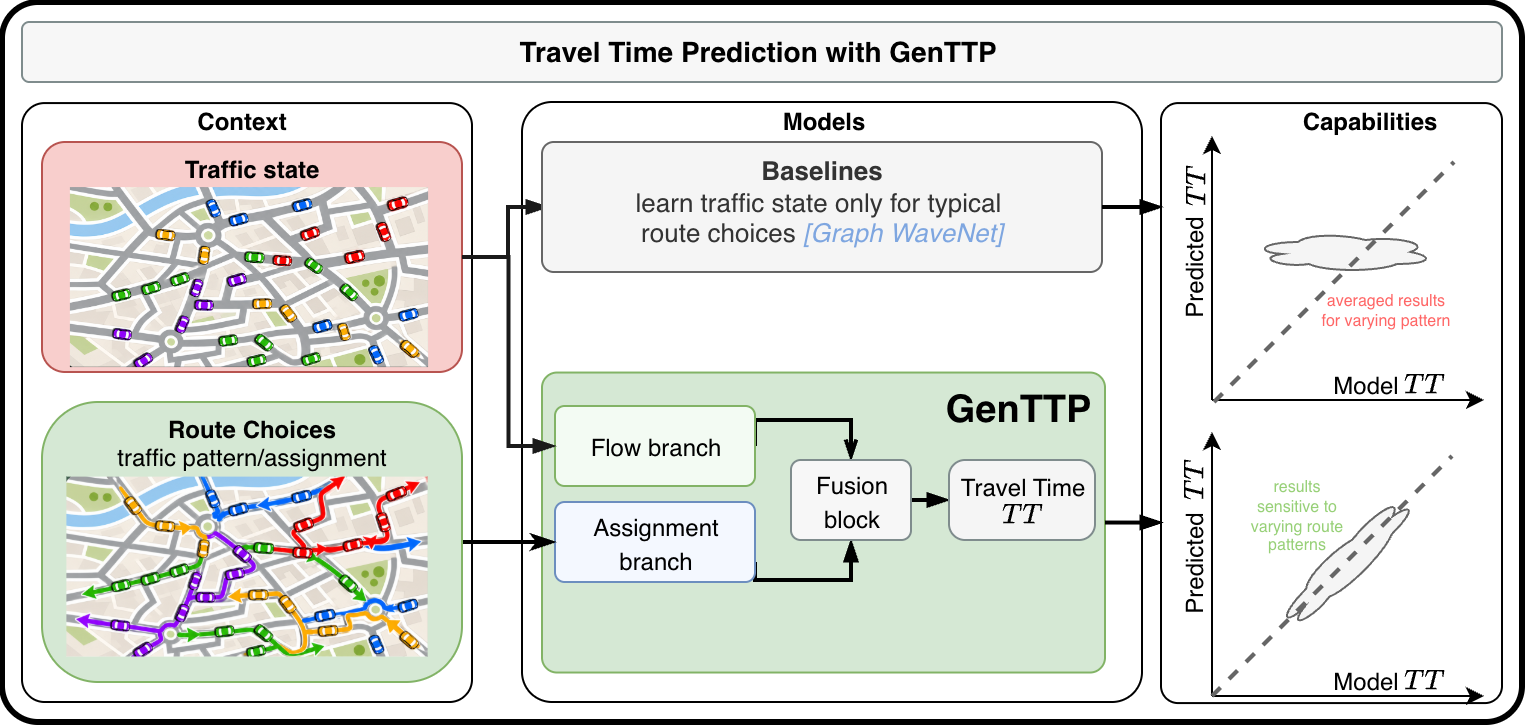}
    \label{fig:teaser}
\end{figure}

We validate our method in real-world urban networks with main results for the city of Inglostadt.
GenTTP provides accurate dynamic estimates for the traffic flows with an average discrepancy of 2.64 per node.
When applied as a recursive method to obtain all future traffic states, we predict the total travel time with only 10,3 \% error, whereas baselines receive from 18\% to 82 \% difference. Computation of \ours inference is faster than microsimulation.

Our contribution can be summarised as follows.
\begin{itemize}
    \item In contrast to existing assignment-agnostic methods, we propose assignment-derived traffic flow predictor conditioned on route choice.
    \item We develop a multi-head neural network that combines microscopic traffic propagation with macroscopic flow directionality.
    \item We propose an assignment processing method that successfully uncovers spatiotemporal complexities arising with varying routing strategies. 
    \item We introduce a Sparsity-Aware Smoothing Mechanism (low-occupancy gate) that explicitly handles the nodes with null traffic (typical for urban traffic), which successfully mitigates the overestimating of MSE-training.%We enhance traffic flow processing algorithm by introducing a low-occupancy gate, which successfully mitigates overestimating tendency of existing methods.
    \item We show that our method provides accurate flow and travel time estimates for varying demand levels in real-world urban networks.
\end{itemize}

\section{Related Work}
\label{sec:related_work}

Existing traffic forecasting models predict future traffic based solely on past traffic states, while \ours also utilises the dictated route assignment.

\paragraph{Graph Neural Networks for Traffic Forecasting.}
Graph neural networks have become a dominant approach for traffic forecasting, where road segments, sensors, or spatial cells are represented as graph nodes and traffic states are modelled as temporal signals over this graph. DCRNN \citep{li2018dcrnn} models traffic evolution as a diffusion process on a directed graph and combines diffusion convolution with recurrent sequence modelling. Graph WaveNet \citep{wu2019graphwavenet} improves this line of work by combining graph convolutions with dilated temporal convolutions and a learned adaptive dependency matrix, allowing the model
to capture both a pre-defined spatial structure and hidden dependencies. A follow-up research showed that Graph WaveNet can be further improved through better hyperparameters, improved gradient flow, short-term pretraining, and ensembling \citep{shleifer2019incrementally}. These models provide strong baselines for spatio-temporal traffic prediction. However, they are designed to forecast future traffic states from past observed traffic states, not to evaluate how different route assignments of the same demand change the resulting network-wide travel time.

\paragraph{Adaptive, Attention-Based, and Transformer Traffic Models.}
Several later models extend traffic forecasting by improving how spatial and temporal dependencies are represented. GMAN \citep{zheng2020gman} uses a graph multi-attention architecture to model relationships between historical and future time steps. AGCRN \citep{bai2020agcrn} removes the need for a fixed pre-defined graph by learning node-adaptive parameters and data-adaptive graph dependencies. STJGCN \citep{zheng2021stjgcn} constructs spatio-temporal joint graphs to explicitly connect multiple nodes across different time steps, rather than only propagating information over a static graph at each time step. Z-GCNETs \citep{chen2021zgcnets} introduce time-aware topological descriptors based on zigzag persistence, while STAEformer \citep{liu2023staeformer} shows that spatio-temporal adaptive embeddings can make a vanilla Transformer \citep{attention_is_all_you_need} competitive for traffic forecasting. These methods improve representation learning for traffic dynamics, but the prediction task remains largely observational: the model receives historical traffic patterns and predicts their continuation. In contrast, \ours is assignment-aware: it conditions the prediction on an explicit route-assignment matrix and is therefore designed for counterfactual evaluation of alternative route-choice patterns.

\paragraph{Learning-Based Traffic Assignment and Assignment Surrogates.}
A related line of work studies learning-based surrogates for traffic assignment. Traditional traffic assignment methods are computationally expensive because they often require iterative optimisation or repeated simulation to approximate equilibrium flows. The recent assignment-focused work of \citet{lassen2025learning} proposes a graph neural network metamodel for approximating stochastic user equilibrium flows, with the goal of accelerating large-scale scenario analysis and improving out-of-distribution assessment. This direction is relatively close in motivation to ours, because both approaches use learning-based models to reduce the computational cost of transportation modelling. However, the prediction target differs. Assignment surrogates typically aim to approximate the outcome of a single assignment procedure, i.e. equilibrium link flows. Contrary to previously mentioned models, the assignment is found, not embedded in the traffic flow structure. Our model instead assumes a candidate assignment as an argument and predicts the resulting traffic evolution and travel time. As such, \ours considers a broader family of efficient assignments, such as a system optimum (with total travel time lower than a user equilibrium) or other assignments that follow particular socioeconomic goals (confer \cite{de2011traffic}).

\paragraph{Generalising Travel Time Prediction to Varying Assignments.}
Existing traffic forecasting models and assignment surrogates leave an important gap. Forecasting models such as DCRNN, Graph WaveNet, GMAN, AGCRN, STJGCN, Z-GCNETs, and STAEformer learn strong spatio-temporal predictors, but they generally assume that the route-choice behaviour behind the observed traffic states is fixed, recurring, or only implicitly represented. Assignment metamodels, on the other hand, focus on approximating assignment outcomes rather than predicting dynamic traffic and travel-time consequences for arbitrary candidate assignments. \ours addresses this gap by learning a mapping from route assignments to traffic flows and system-wide travel time. Given the same travel demand but different route assignments, the model is expected to distinguish the resulting congestion propagation and predict different travel-time outcomes. This makes \ours particularly relevant for coordinated fleets and connected autonomous vehicles, where route choices are not merely observed but actively controlled.

\section{Preliminaries}
\label{sec:preliminaries}

\paragraph{Route assignments in traffic networks.}
Urban traffic performance is determined not only by the amount of demand (number of vehicles trips), but also by how this demand is distributed over the available routes. Let \(D\) denote a fixed set of trips, where each vehicle has an origin, a destination, and a departure time. For every vehicle, a finite set of feasible network paths is assumed to be available. A route assignment specifies which path is selected by each vehicle. If \(Q\) vehicles each choose from \(K\) candidate paths, the number of possible assignments grows as \(K^Q\). This combinatorial growth makes exhaustive evaluation infeasible even for moderately sized networks. The central difficulty addressed in this paper is that two assignments with the same demand \(D\) may produce substantially different congestion patterns and, therefore, different system-wide travel times - as we demonstrate on our experiments.

\paragraph{Microscopic simulation as a reference model.}
Microscopic traffic simulators model the movement and interaction of individual vehicles over time. Given a road network, a travel demand, and a route assignment, simulators such as SUMO, VISSIM, or Aimsun produces vehicle trajectories, realised flows, and travel times \citep{Vissim, aimsun}. These simulators are accurate and flexible because they account for local interactions between vehicles and for the propagation of congestion through the network. However, they are computationally expensive when many alternative route assignments must be evaluated. This limitation is particularly important in assignment optimisation, where a routing policy may require thousands or millions of candidate assignments to be assessed.

\paragraph{Traffic load representation.}
To use microscopic simulation outputs in a learning framework, vehicle trajectories are aggregated into a spatio-temporal traffic load representation. The road network graph is aggregated into \(S\) spatial units, and the \(i\)-th simulation horizon is divided into \(T_i\) time intervals. The resulting traffic load matrices are denoted by
\[
Q(i)\in \mathbb{N}^{S \times T_i},
\]
where each entry represents the number of vehicles observed in a given spatial unit at a given time step. This representation is commonly used by spatio-temporal graph learning models as \cite{WaveNet}, which treat traffic as a dynamic signal evolving over a graph.

\paragraph{Assignment representation.}
In this work, route choices are also represented as a spatio-temporal matrix. The selected path of each vehicle is mapped to the spatial units it traverses and to the time interval in which it departs or contributes to expected load. This yields assignment matrices
\[
A(i) \in \mathbb{N}^{S \times T_i},
\]
which summarises how planned route choices distribute vehicles over the network. Unlike the realised traffic load \(Q({i})\), which is obtained after running the simulator, \(A({i})\) is known before simulation. It therefore provides direct information about the routing decision whose consequences we want to predict.

\paragraph{From traffic forecasting to assignment-conditioned prediction.}
Most graph-based traffic forecasting models learn to predict future traffic states from past observed states. In simplified form, they estimate
\[
(Q_{t-W_Q}, \ldots, Q_{t-1}) \mapsto \hat{Q}_t,
\]
where \(W_Q\) is a historical observation window. This setting is suitable when traffic follows typical recurring patterns, but it does not explicitly encode alternative route assignments. Consequently, such models are not designed to answer an alternative assignment question: what would happen if the same vehicles departed at the same times but selected different routes?

\ours considers a different prediction setting. The model receives both historical traffic load and route-assignment information, and predicts the traffic state induced by that assignment:
\[
(A_{t-W_A}, \ldots, A_{t-1}, Q_{t-W_Q}, \ldots, Q_{t-1}) \mapsto \hat{Q_t}.
\]
The predicted flow sequence \(\hat{Q}=(\hat{Q}_{W_Q}, \hat{Q}_{W_Q+1}, \ldots, \hat{Q}_{T})\) is then aggregated into the system-wide travel time \(TT\). This makes the model a surrogate for microscopic simulation: instead of repeatedly running a computationally costly simulator for every candidate route assignment, GenTTP learns to approximate the simulator response to varying assignments.

By conditioning the prediction on an explicit route-assignment matrix, \ours moves beyond simple pattern continuation. This allows the model to answer interventional questions regarding traffic conditions resulting from alternative demand distribution across paths, a capability absent in state-of-the-art.

\section{GenTTP}
The primary objective of \ours is to develop a robust mechanism for predicting the system-wide travel time in environments where vehicle routes are the control variable\footnote{The code for our model is available at \url{https://anonymous.4open.science/r/genttp/}.}. The proposed approach accounts for how spatiotemporal fluctuations in traffic density and changing routing strategies influence overall network performance.

\subsection{Problem Statement}

Let us consider an arbitrary traffic model $SIM$ that with desired realism simulates the propagation of traffic demand $D$ along their chosen paths $A$ across some network $S$ and allows to reliably estimate the resulting total travel times $TT$. The demand $D$ is, for now, assumed fixed, i.e. in every simulation exactly the same drivers depart from their origins to their destinations at the same departure times and changes occur only in agents' route assignments. We assume that each agent may freely choose a path from her own precomputed choice set (paths sampled from \texttt{JanuX} \cite{JanuX}), yielding an enormous search space of $K^Q$ possible assignments, quickly exploding with both route choice-set $K$ and the number of agents $Q$.

\ours~is a surrogate model trained to approximate the realistic simulator $SIM$. In our case, it~approximates the SUMO \citep{SUMO} microscopic traffic flow model. Is is applied to predict how 1034 agents traversing 734 nodes of the Ingolstadt network commute within a day scenario. It first collects the offline precomputed training data, i.e. multiple independent simulation runs for various route choices (assignments) $A \in \mathcal{A}$, where \(\mathcal{A}\) is a set of admissible routing patterns. Specifically, for the \(j\)-th simulation (\(j \in \{1, \ldots, K\}\), where \(K\) is a number of recorded simulations), we consider assignment \(A(j) \in \mathcal{A}\) on the network $\mathcal{N}$ with the fixed demand $D$. The simulation model yields, for different $A(j)$, respective traffic flows $Q_t(j)$ and travel times $TT(A(j))$. The objective of \ours is to then approximate the microscopic model as closely as possible: $\text{\ours} \approx SIM$.

\subsection{Model Formalisation}
\ours maps route assignment to an estimate of the aggregate system-wide travel time:
\begin{equation}
GenTTP:\quad \mathcal{A} \ni A \mapsto TT \in \mathbb{R}_{+}.
\end{equation}

One simulation is associate with a single route assignment $A \in \mathcal{A}$, formulate as $A \in \mathbb{N}^{S \times T}$, where $S$ and $T$ represent spatial and temporal resolutions, respectively.
% For \(j\)-th simulation, $A \in \mathcal{A}$ represents the drivers' route assignment, formalised  as $A \in \mathbb{N}^{S \times T_j}$, where $S$ and $T=T_j$ represent spatiotemporal resolution of the problem (vehicles propagate among $S$ cells in time-steps $T$). 
% The natural input is a vector of paths selected by each agent (sequence of nodes) and their departure times (fixed). In \ours, we map the selected paths and the departure time to the assignment matrix $A$. 
For each agent, we represent their assignment as a binary vector which marks nodes on the path. 
Then, for each time step, we aggregate the information.
To formalise, each column of assignment matrix A, i.e. $A_t$, where $t~\in~\{1, \ldots, T\}$, is given by a vector $A_t = \sum_{i=1}^{v_t} A_{t}^{i}$ for $v_t$ - the number of vehicles departing in the time period $t$ and $A_{t}^{i}$ - a binary vector such that $(A_{t}^{i})_s = 1$ if the $s$-th node is on the path of vehicle $i$ and $(A_{t}^{i})_s = 0$ otherwise. Such representation proves to be both computationally efficient and meaningful for the prediction problem.

The implicit intermediate representation is the traffic flow states $Q = Q(SIM(A))$, typically used in similar approaches \citep{WaveNet}. It is derived from the simulator result as $Q \in \mathbb{N}^{S \times T_i}$ with each entry in the $Q$ matrix representing the vehicle flow observed in a given space (node) in a given time interval. We use $\hat{Q}$ to represents approximates given by our model.

The network-wide travel time $TT \in \mathbb{R}_{+}$ is a scalar output, which we derive from $\hat{Q}_t$ by an aggregation function $g: \hat{Q} \to TT  \in \mathbb{R}_{+}$. Focusing on the accuracy of the predicted flow matrix $\hat{Q}$, \ours captures the underlying spatiotemporal dynamics of traffic flow that drive the resulting travel times $TT$. 
Since $\hat{Q}_t$ represents the number of vehicles present in each spatial cell at time $t$, the aggregated travel time can be approximated as $g(\hat{Q}) = \sum_{t=1}^{T-1} \delta_t \lVert \hat{Q}_t \rVert$, where $\delta_t$ refers to a time step length and $\lVert \hat{Q}_t\rVert$ denotes the total flow vector, i.e. $\lVert \hat{Q}_t\rVert = \sum_{i=1}^S \hat{q}_i(t)$, where $\hat{q}_i(t) = (\hat{Q})_{t,i}$.
Despite the non-linear relationship between traffic flow and travel times, the proposed proxy proves to efficiently estimate system-level performance with short time intervals \(t\).

\subsection{Model architecture}

\begin{figure}[hbp]
    \centering
    \includegraphics[trim=3cm 4cm 0cm 1cm, scale=0.35]{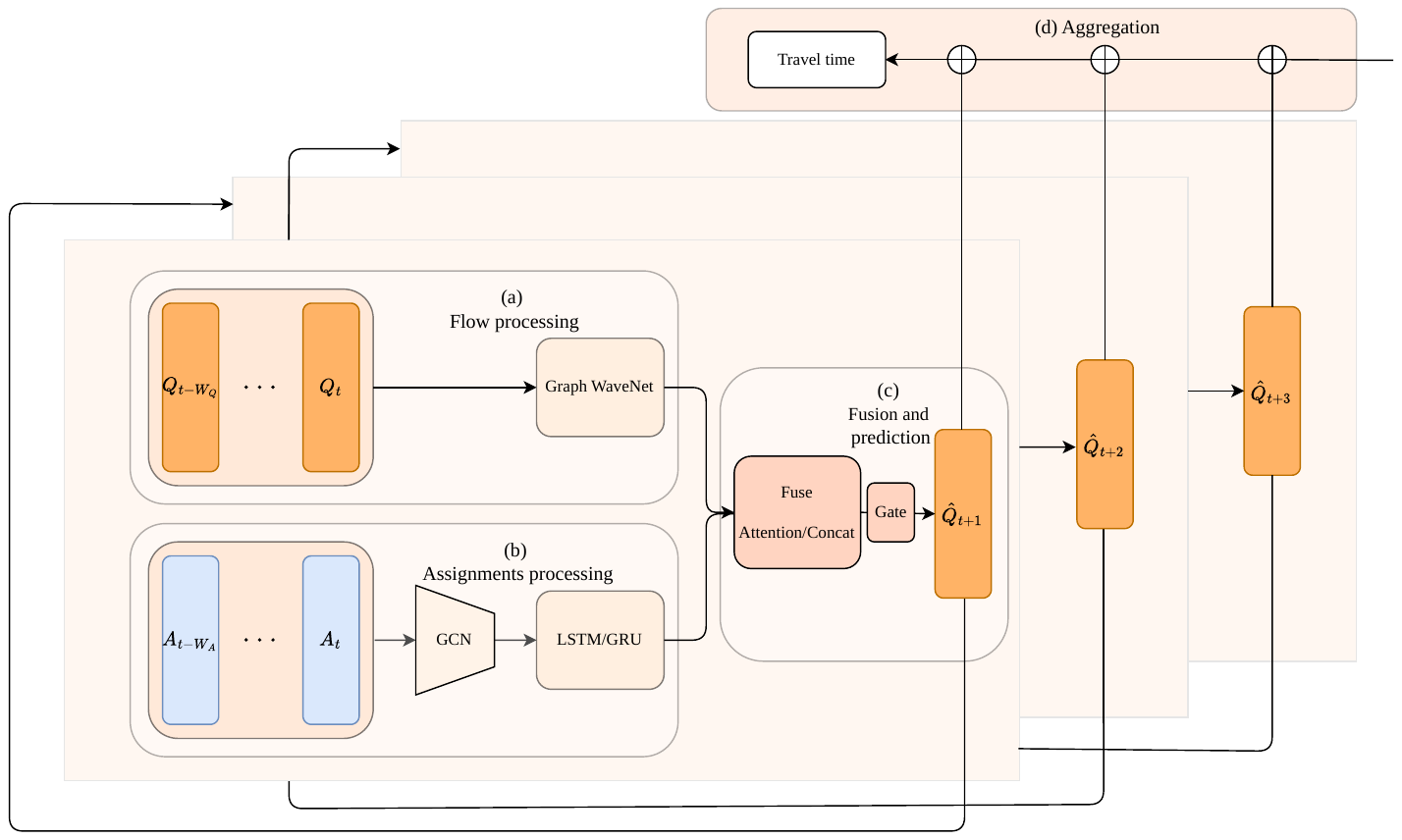}
    \caption{Architecture of \ours. 
    \textbf{(a) Flow processing branch}: spatio-temporal encoding of previous traffic states using a Graph WaveNet-based GCN. 
    \textbf{(b) Assignment processing branch}: time sequence modelling of planned path assignments using recurrent units (LSTM/GRU) to capture future network load.
    \textbf{(c) Fusion and prediction}: attention-based/concatenation mixture of both branches followed by an MLP decoder and sigmoid gate to predict the next-state flow $\hat{Q}_t$, which are then aggregated to estimate the system-wide travel time $TT$.
    \textbf{(d) Aggregation}: calculation of the total travel time with the aggregation function on the approximated flows.}
    \label{fig:adttp_architecture}
\end{figure}

The proposed \ours architecture follows a deep learning approach based on neural network to capture both local correlations of assignment matrices and global impact of departing vehicles.
We use WaveNet \citep{WaveNet}, a Graph Convolutional Network (GCN) \citep{gcnn} that relies on a predefined adjacency matrix between spatial nodes $S$ to process flow matrices \(Q\) (Fig.~\ref{fig:adttp_architecture}(a)). The processing of the assignments $A$ is conducted by the dedicated sequence modelling branch. The proposed architecture applies Long Short Term Memory (LSTM) and Gated Recurrent Unit (GRU), but is not limited to these choices. By using the high-dimensional representation of assignments, we capture the incoming traffic flow (Fig.~\ref{fig:adttp_architecture}(b)).
Data processed on independent branches are finally fused and then passed through a Multi Layer Perceptron (MLP) network (Fig.~\ref{fig:adttp_architecture}(c)). 
For the data fusion, framework serves attention-based and data concatenation modules.
Then MLP supported with skip-connection architecture maps the fused feature vector to the next-step flow prediction scaled via sigmoid gate.
Lastly, we apply an aggregating function to the approximated flows to obtain the total travel time (Fig.~\ref{fig:adttp_architecture}(d)).

The units depicted in Fig.~\ref{fig:adttp_architecture}(a)-(c) allow us to predict the next-step flow $\hat{Q}_t$ based on previous traffic states $\hat{Q}_{\tau: \tau < t}$ and path assignments $A$. By iterating this process, \ours sequentially predicts the flow matrices for all subsequent time steps, which are then aggregated to estimate the system-wide travel time $TT$ (Fig.~\ref{fig:adttp_architecture}(d)). Consequently, \ours predicts the complete traffic flow matrix $\hat{Q} \in \mathbb{N}^{S \times T}$ expected to result from a given assignment $A$. 

The recurrent part of the model (Fig.~\ref{fig:adttp_architecture}(a)-(c)), denoted by $f_t$, utilises a moving window approach for the temporal feature extraction. Specifically, the input sequence consists of the $W_Q$ historical traffic load states and the $W_A$ historical assignment. Formally, for any time step $t$ such that $t \in \{\max\{W_{Q}, W_A\} + 1, \ldots, T \}$, we define functions
$$ f_t: \quad \mathcal{A}^{W_A} \times \mathcal{Q}^{W_Q} \ni (A_{t- W_{A}}, A_{t - W_{A} + 1}, \ldots, A_{t-1}, Q_{t-W_Q}, Q_{t-W_{Q} + 1}, \ldots, Q_{t-1}) \mapsto \hat{Q}_t \in \mathbb{R}_{+}^{S},$$
where indexes denote the respective columns of $A, Q$ and $\hat{Q}$.

\subsection{GenTTP Model Training}

\ours is trained on a dataset of traffic simulations exploring the relation between assignments $A$ and travel times as broadly as possible. We note that the search space quickly becomes too big for an exact search (as even for 20 vehicles selecting from 5 paths there is $5^{20}$ possible assignments); thus, a smart sampling is crucial to train the model to reliably predict travel times.

To ensure robust model performance and prevent overfitting, the dataset was split into train, test, and validation subsets, according to the 70/10/20 respective ratio. Training was conducted using mini-batches of size 128.
Parallelisation is achieved through GPU-accelerated tensor operations, particularly, using PyTorch einsum kernels for graph convolutions across all 195 spatial nodes simultaneously. 

% While the proposed method is applicable for any spatial and temporal resolutions (e.g. propagation across detailed microscopic road segments updated every tenth of the second), we anticipate computation issues.

The grid-based sampling strategy on the training data explicitly exposes the model to OOD (Out-of-Distribution) congestion patterns that are rarely present in standard historical datasets.

\section{Experiments}

We demonstrate \ours to predict the travel time on the Ingolstadt network (from \cite{intas}), with 1034 vehicles traversing the network of 734 nodes. We trained it with 1 000 microscopic simulations, where we simulated how total travel time changes for various demand patterns (for training details, refer to Appendix \ref{app:hyperparameters_training}). In every simulation, the set of vehicle departures and origin-destination pairs remains fixed, while only the route selected for each agent is varied (detailed in Appendix \ref{app:sumo}). 
To efficiently train the model we used grid-sampling method (explained in Appendix \ref{app:janux}) intended to cover a wide range of assignment patterns.
We apply spatial aggregation using the H3 hexagonal indexing system \cite{H3}. Temporal aggregation is performed at a 10-second resolution. Overall the dataset translates to 297195 individual samples

Although the objective of \ours is to minimise the absolute error in predicting the aggregate system travel time ($\Delta TT$), we track its precision by comparing the predicted flow in intermediate steps to the actual flow. 
This is successfully captured by the absolute metrics MAE and RMSE. 

\subsection{Models Comparison}
For baselines, we choose other GNN traffic predictors: STAEFormer \citep{liu2023staeformer} and ADGCRN \citep{duan2023learning}.
We also conduct the ablation study of our model by comparing it to pure flow-based branch (model WaveNet \citep{WaveNet}) and with various architectural options.
The results in Table~\ref{tab:baselines} demonstrate that including the assignment matrix $A$ improves prediction accuracy compared to purely flow-based models, confirming that using assignment as a direct input is a critical component for surrogate modelling in coordinated CAV environments.
\begin{table}[h t b p]
\centering
\caption{Performance comparison against Baseline models: STAEformer, ADGCRN and Graph WaveNet with \ours framework for both Concatenation and Attention-based fusion module.}
\label{tab:baselines}
\begin{tabular}{lcccc}
\hline
\textbf{Method} & \textbf{MAE} & \textbf{RMSE} & \textbf{TT difference [min] } & \textbf{Relative TT difference}\\ \hline
STAEformer & 2.64 & 6.55 & 9301 & 18.7 \% \\
\hline
ADGCRN & 2.71 & 6.63 & 36786 & 76 \% \\
\hline
Graph WaveNet & 2.7 & 6.58 & 39956 & 82.7 \% \\
\hline
{GenTTP (Concat)} & \textbf{2.52} & \textbf{6.05} & 5852 & 12 \% \\
{GenTTP (Attention)} & 2.64 & 6.45 & \textbf{4991}  & \textbf{10.3 \%} \\ \hline
\end{tabular}
\end{table}

\begin{figure}[p]
    \centering
    \caption{Comparison of \ours with baseline models on the task of predicting system-wide travel time. 
     Figure~\ref{fig:assignment_wavenet} shows the results obtained with Graph WaveNet, whose predictions remain largely uncorrelated with the true travel times. The results are averaged per assignment, and even small temporal discrepancies in travel time can mislead the prediction. Figure~\ref{fig:assignment_staeformer} presents the STAEformer architecture, which was able to produce varying travel time estimates, but without a meaningful correlation with the ground-truth travel times.
     Figure~\ref{fig:assignment_fourth_model} presents \ours with concatenation-based fusion. This variant is able to distinguish between different assignments, but it tends to underestimate more congested simulations, likely because the traffic-state input has a weaker influence in the fusion process. Finally, Figure~\ref{fig:assignment_lstm} shows the performance of \ours with an attention-based fusion module. The model learns to assign higher travel times to assignments that induce stronger overlap in demand patterns.}
    \begin{subfigure}[t]{0.49\textwidth}
        \centering
        \includegraphics[width=\linewidth]{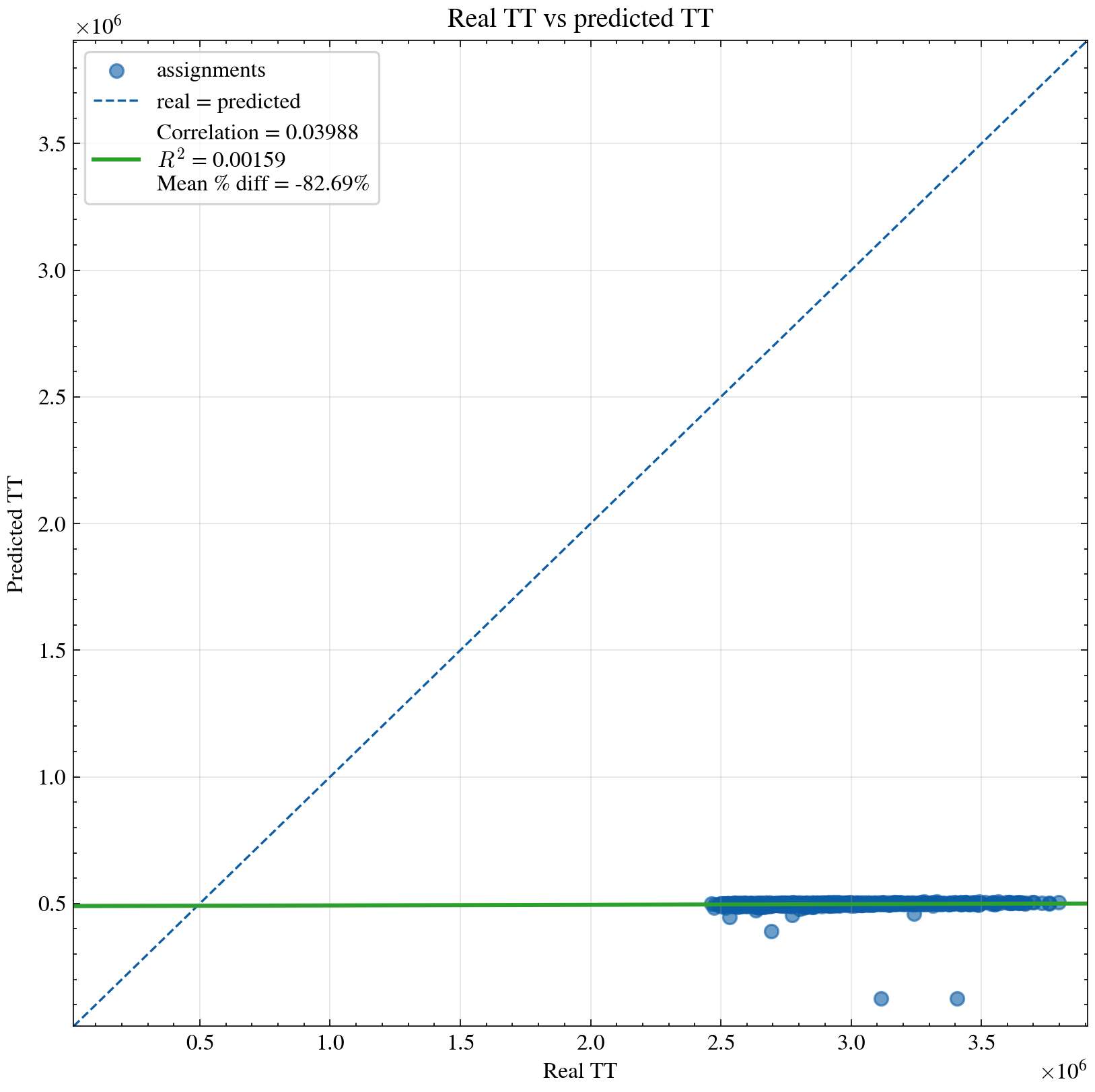}
        \caption{Graph WaveNet}
        \label{fig:assignment_wavenet}
    \end{subfigure}
    \hfill
    \begin{subfigure}[t]{0.49\textwidth}
        \centering
        \includegraphics[width=\linewidth]{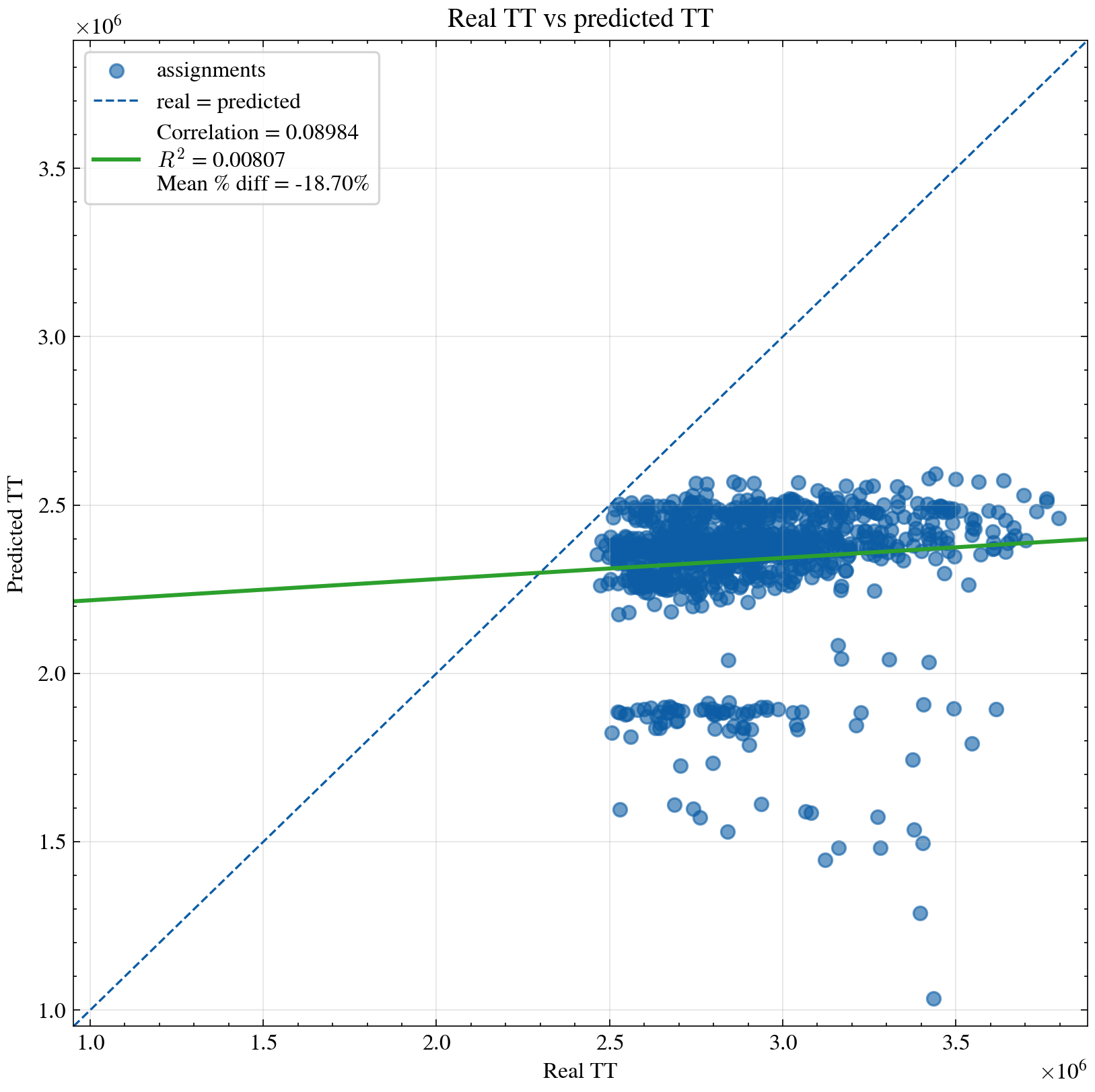}
        \caption{STAEformer}
        \label{fig:assignment_staeformer}
    \end{subfigure}

    \vspace{0.75em}

    \begin{subfigure}[t]{0.49\textwidth}
        \centering
        \includegraphics[width=\linewidth]{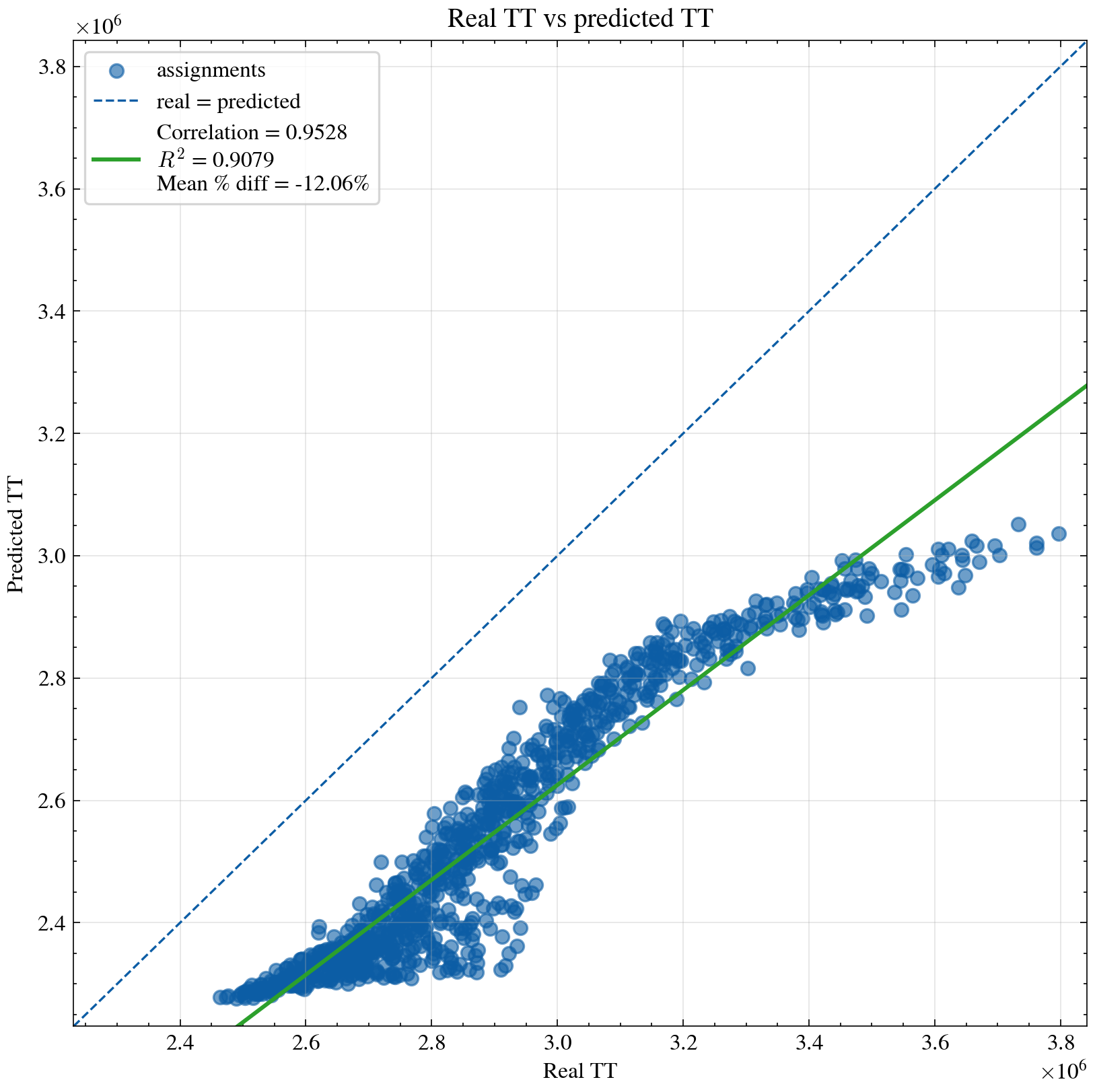}
        \caption{GenTTP Concatenate}
        \label{fig:assignment_fourth_model}
    \end{subfigure}
    \hfill
    \begin{subfigure}[t]{0.49\textwidth}
        \centering
        \includegraphics[width=\linewidth]{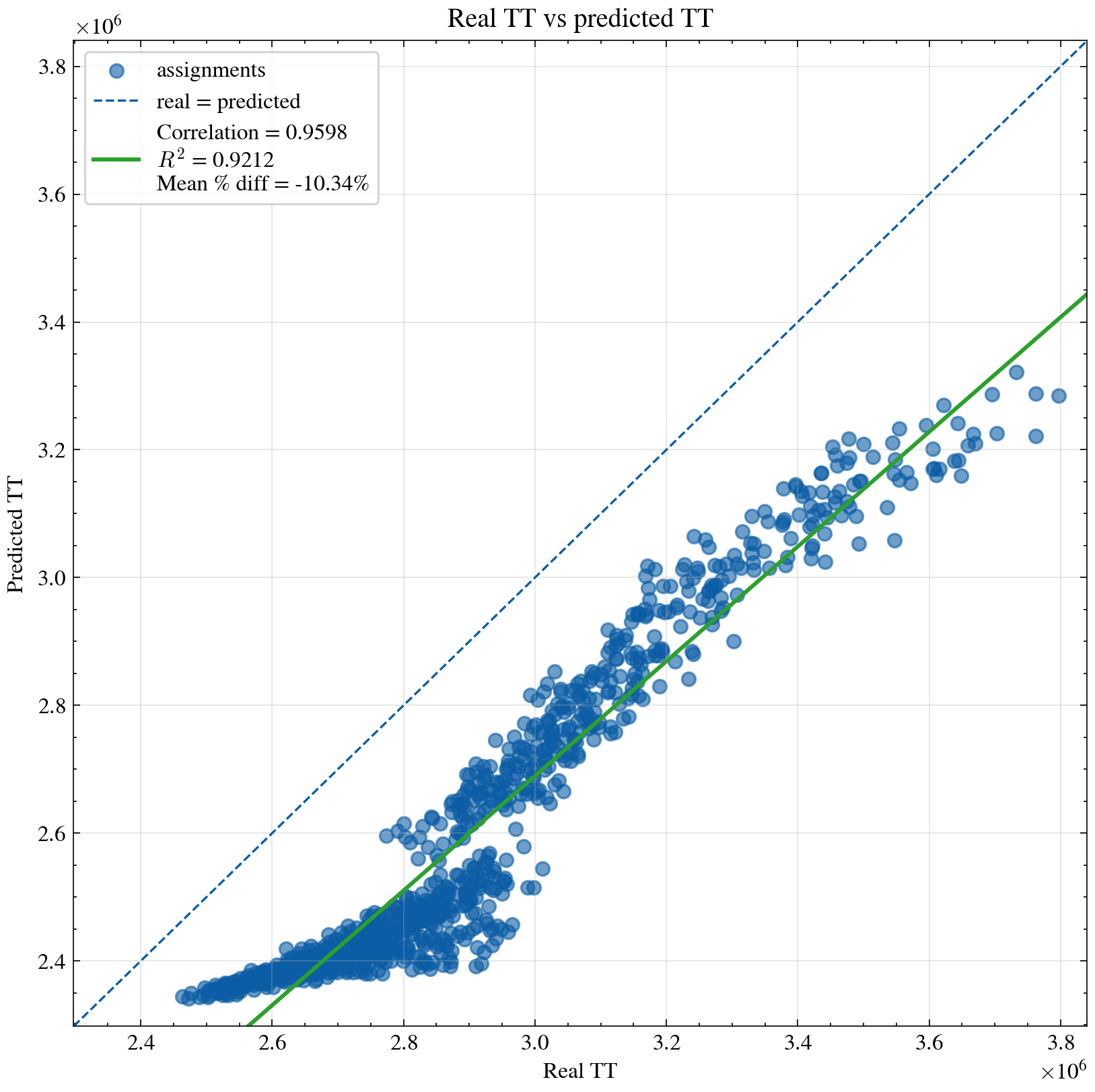}
        \caption{GenTTP Attention}
        \label{fig:assignment_lstm}
    \end{subfigure}
    \label{fig:assignment_branch_impact}
\end{figure}

\subsection{Node-wise prediction}

\begin{figure}[ht]
    \caption{Comparison of \ours and baseline (STAEformer) results in node-wise flow prediction across time steps. Figure~\ref{fig:crowded_node} shows predictions for a spatial cell with high average flow, while Figure~\ref{fig:not_crowded_node} shows predictions for a less frequently used cell.}
    \label{fig:flow_nodes}
    \centering
    \begin{subfigure}[t]{0.49\linewidth}
        \centering
        \includegraphics[width=\linewidth]{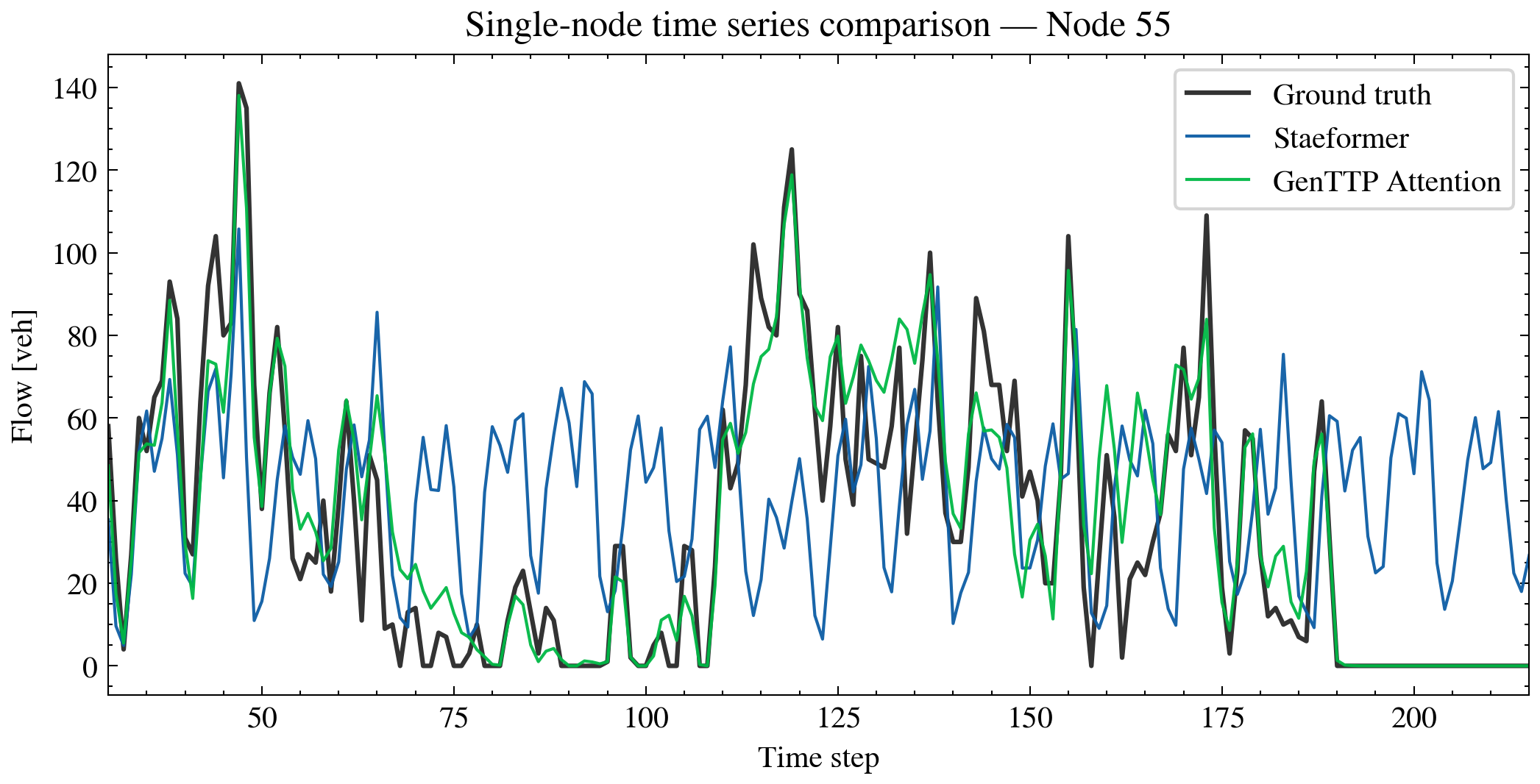}
        \caption{Prediction for a highly congested spatial cell.}
        \label{fig:crowded_node}
    \end{subfigure}
    \hfill
    \begin{subfigure}[t]{0.49\textwidth}
        \centering
        \includegraphics[width=\linewidth]{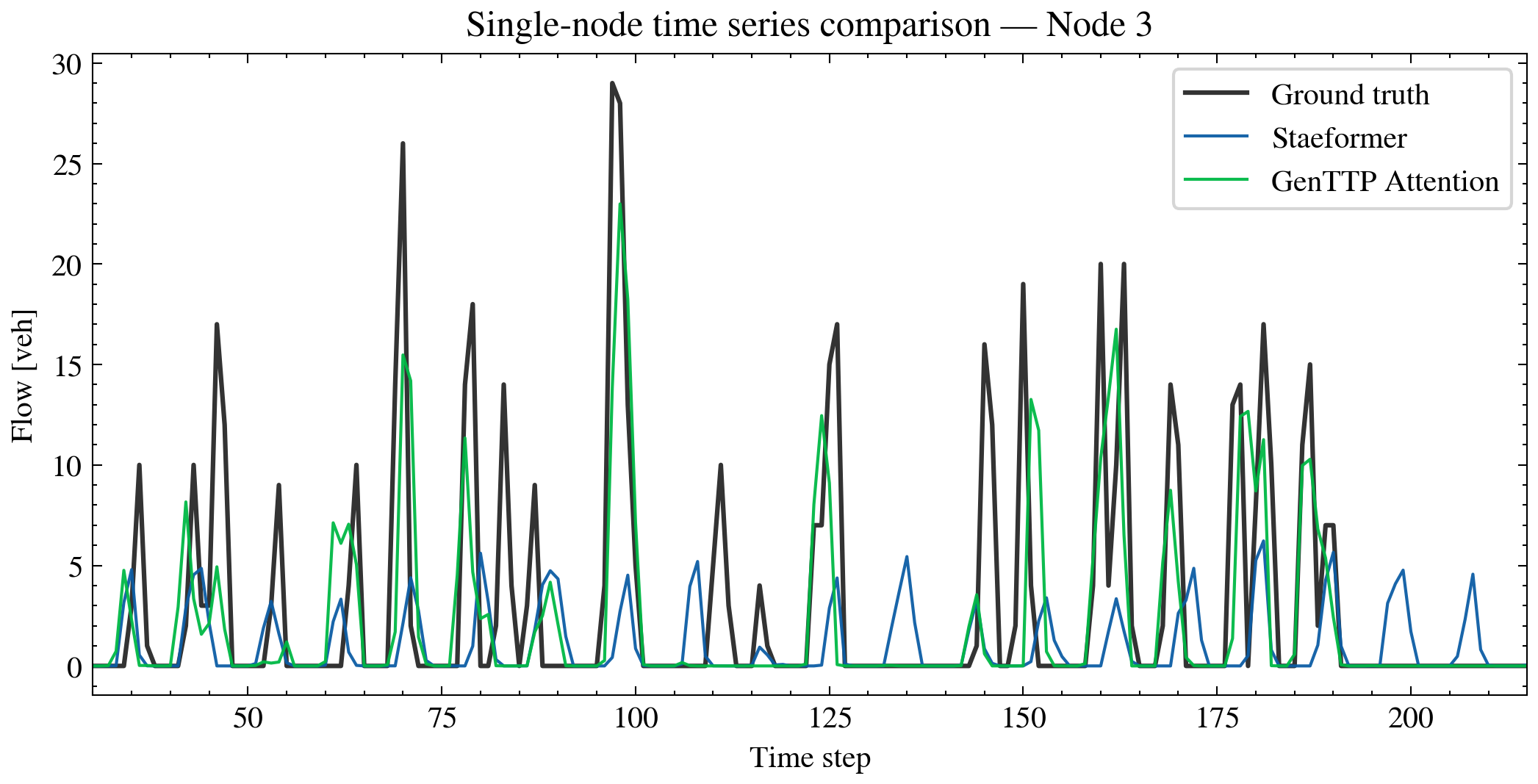}
        \caption{Prediction for a less frequently used spatial cell.}
        \label{fig:not_crowded_node}
    \end{subfigure}
\end{figure}

Figure \ref{fig:flow_nodes} shows that our model predicts traffic flow on particular nodes well. On a densely congested node (Fig. \ref{fig:crowded_node}), \ours follows trends induced by the particular assignment. Conversely, we note that flow-based models fail to capture dynamic changes in the occupancy. In Figure \ref{fig:not_crowded_node}, we analyse a sparsely populated node. Our gated approach successfully recognises nullified congestion and limits over- and underestimation.

\section{Conclusion and Discussion}

We introduced \ours that generalises urban travel time predictions to varying route choices. It is a learning-based framework for estimating system-wide travel time in urban traffic systems under varying route assignments. In contrast to previous predictive approaches that rely on observed traffic states, the proposed method explicitly incorporates vehicle route-choice information, making it suitable for settings in which routing decisions vary across scenarios.

The proposed framework extends graph-based deep learning for traffic modelling to the assignment-aware setting and offers a computationally efficient surrogate for repeated traffic simulation. By combining aggregated traffic representations with structured assignment inputs, \ours captures how routing decisions influence congestion propagation and, in turn, total travel time. 

The results show that this combination enables efficient prediction of network-level travel time while significantly reducing the computational burden associated with microscopic simulation.

These findings highlight the potential of assignment-aware predictive models as tools for rapid scenario evaluation and decision support, particularly in applications involving coordinated fleets of connected and autonomous vehicles. More broadly, the proposed framework provides a basis for learning-based surrogate models for dynamic assignment problems.

However, our results are currently confined to a specific micro-simulator and limited demand variety. Therefore, they need to be further validated to assess the model's generalisability.

%\section{Acknowledgements}

\bibliographystyle{plainnat}
\bibliography{references}

\appendix

\newpage

\appendix

\section{Datasets}
\label{app:datasets}

This appendix provides additional details on the datasets used to train and evaluate \ours. Our main experiments are conducted on the Ingolstadt SUMO network introduced in \cite{intas}. In this setting, a fixed travel demand pattern is simulated under varying route assignments. Each simulation corresponds to a single assignment realization \(A\), from which we extract the resulting traffic load matrix \(Q\) and the aggregate travel time \(TT\).

\subsection{SUMO Simulator}
\label{app:sumo}

The dataset contains 1000 SUMO simulations. In every simulation, the set of vehicle departures and origin--destination pairs remains fixed, while only the route selected for each agent is varied. This setup allows the learning problem to isolate the effect of route-choice assignments on congestion formation and total travel time at the network level.

Because microscopic simulation snapshots can produce a very large number of vehicle-level states, we apply spatial aggregation using the H3 hexagonal indexing system \citep{H3}. Temporal aggregation is performed at a 10-second resolution. This choice preserves important traffic dynamics, such as traffic light cycles and the use of frontage roads, while keeping the data representation computationally manageable. Parameters of a chosen network are described in \ref{tab:appendix_dataset_overview} and the SUMO parameters are depicted in \ref{tab:appendix_sumo_config}.

\begin{table}[h]
\centering
\caption{Parameters of Ingolstadt traffic network used in simulations and gathering data details.}
\label{tab:appendix_dataset_overview}
\begin{tabular}{ll}
\toprule
Property & Value \\
\midrule
Network & Ingolstadt SUMO subnetwork \\
Number of agents & 1,034 \\
Number of network edges & 734 \\
Spatial aggregation & 195 spatial cells \\
Temporal resolution & 10 seconds \\
Simulation horizon & from 35 to 65 minutes \\
Number of simulations & 1000 \\
\bottomrule
\end{tabular}
\end{table}

\begin{figure}[h]
\centering
\caption{Spatial aggregation of Ingolstadt network into 195 cells. On the figure hexagons are there placed on region of Ingolstadt city previewed via \cite{OpenStreetMap}.}
\includegraphics[width=0.9\linewidth]{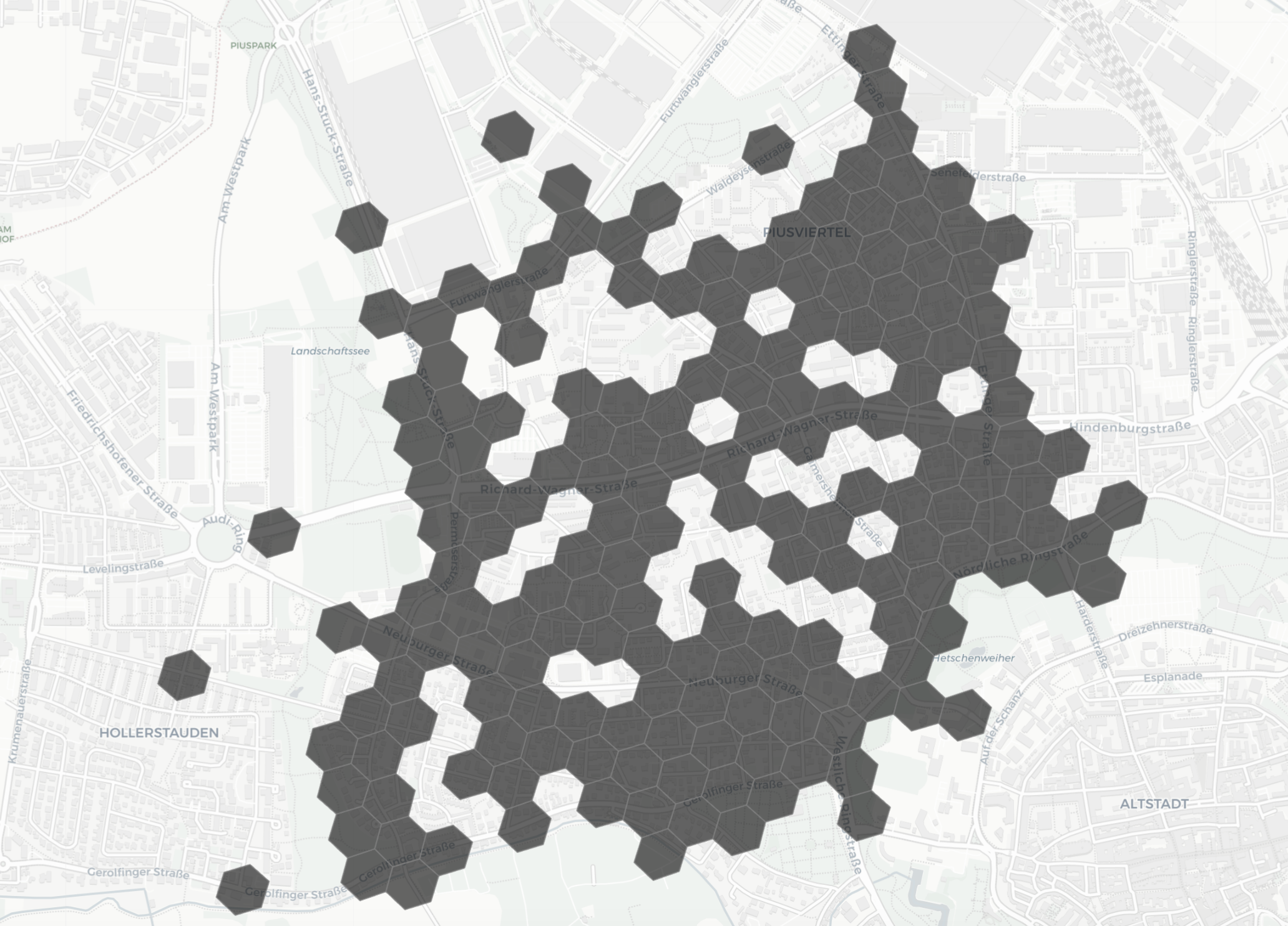}
\label{fig:appendix_ingolstadt_network}
\end{figure}

For trip demand, we use data from \cite{resco} with 306 unique origin destination pairs and departure times spanned within first 30 minutes part of simulations. That demand firstly appeared as a part of benchmark for traffic-light control. 

% SUMO is used as the microscopic traffic simulator that provides reference traffic flows and travel times. In the deterministic setting, repeated simulations with identical route assignments produce identical travel times. This property is useful for constructing a controlled supervised-learning dataset, because variation in the target values can be attributed to variation in route assignments rather than simulator stochasticity.

% For each assignment \(A\), SUMO returns the realised vehicle trajectories and time-dependent traffic states. These outputs are aggregated into the traffic load matrix \(Q \in \mathbb{N}^{S \times T}\), where \(S\) denotes the number of spatial cells and \(T\) denotes the number of time steps. The aggregate travel time \(TT\) is computed from the simulated vehicle trajectories and is used as the final system-level performance measure.

\begin{table}[h]
\centering
\caption{SUMO simulation configuration.}
\label{tab:appendix_sumo_config}
\begin{tabular}{ll}
\toprule
Parameter & Value \\
\midrule
SUMO version & 1.26.0 \\
Random seed policy & Deterministic unless explicitly stated \\
Outputs used & Node flows, travel times, departure times \\
\bottomrule
\end{tabular}
\end{table}

\begin{figure}[h]
\caption{Screenshot of a simulated network state displayed in the SUMO graphical user interface. Each triangle represents a vehicle, with its apex indicating the direction of travel. In this visualization, a custom color overlay was applied so that vehicles with similar origin--destination pairs and departure times are shown in similar colors.}
\centering
\includegraphics[width=0.9\linewidth]{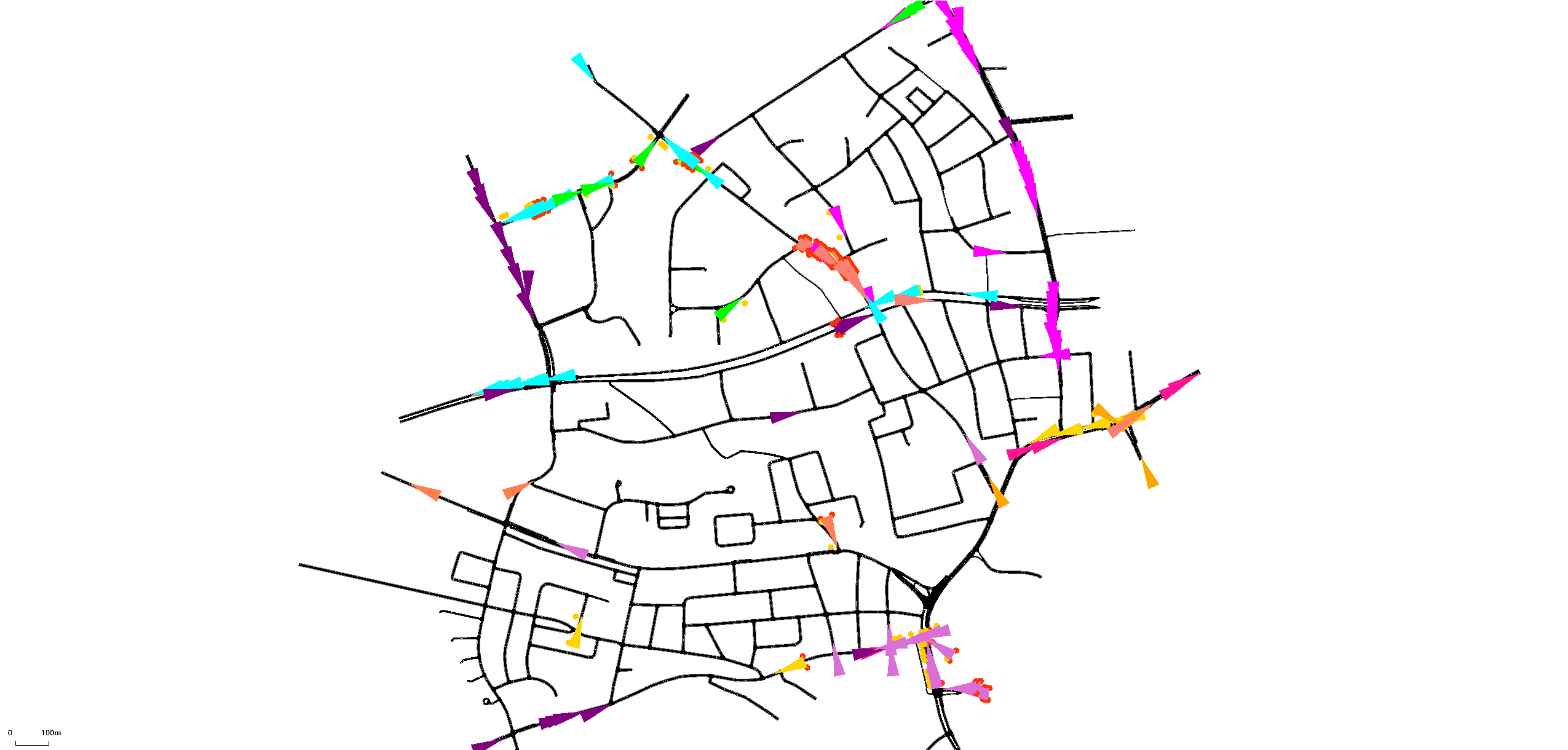}
\label{fig:appendix_sumo_snapshot}
\end{figure}

\subsection{JanuX}
\label{app:janux}

JanuX is used to construct route choice sets for each origin--destination pair. For each agent, a finite set of feasible candidate paths is precomputed before simulation. During data generation, the assignment sampler selects one path from the available choice set for each agent. This produces a complete assignment vector, which is then mapped to the assignment matrix \(A\). The route choices specifics are depicted in \ref{tab:appendix_janux}

The use of precomputed route choice sets has two practical advantages. First, it restricts the otherwise intractable route-choice space to a manageable set of feasible alternatives. Second, it allows the training data to systematically explore different route-choice combinations while preserving physically meaningful paths.

\begin{table}[h]
\centering
\caption{Route choice-set construction summary.}
\label{tab:appendix_janux}
\begin{tabular}{ll}
\toprule
Property & Value \\
\midrule
Path generation tool & JanuX \\
Choice set size per OD pair & \(4\) \\
Number of agents & 1034 \\
Number of unique OD pairs & 306 \\
Output format & Candidate route list and action mask \\
\bottomrule
\end{tabular}
\end{table}

\begin{figure}[!ht]
\centering
\caption{Example route choice set generated for one OD pair. It tends to generate unique paths with some shared road segments in its.}
\includegraphics[width=0.8\linewidth]{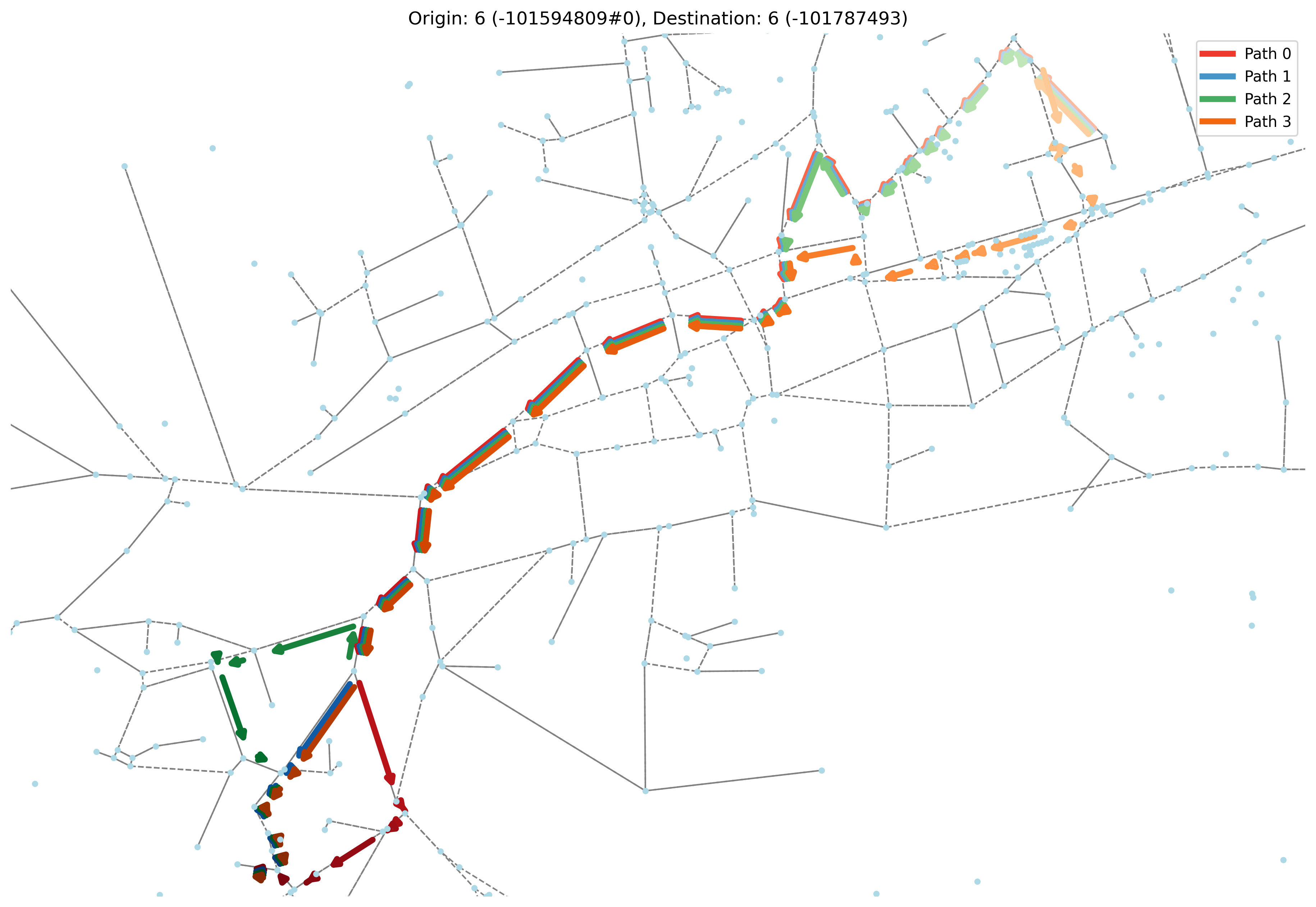}
\label{fig:appendix_janux_paths}
\end{figure}

\begin{figure}[ht]
\caption{The grid sampling method generates assignments by applying evenly spaced changes to the probability vector over the space of feasible path masks.}
\centering
\includegraphics[width=0.6\linewidth, trim = {5cm 14cm 0 0},clip]{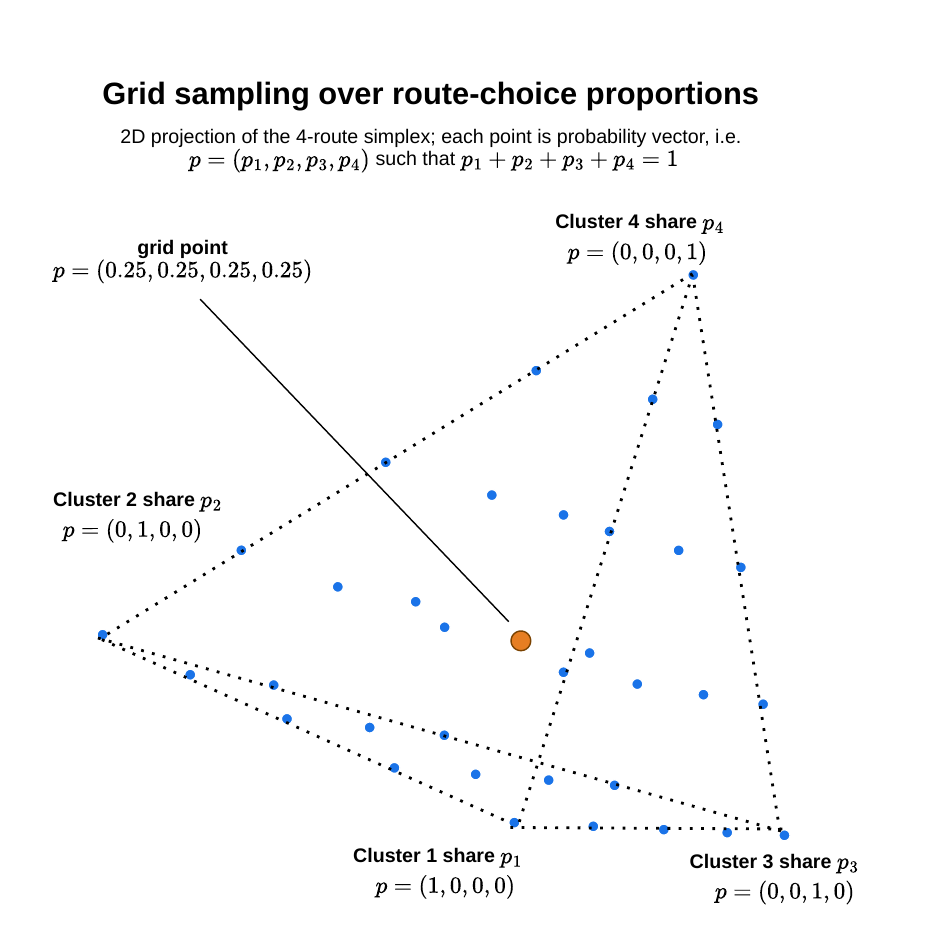}
\label{fig:appendix_grid_simplex}
\end{figure}

The grid method is used to generate route-assignment samples in a structured way. Instead of sampling each assignment fully at random, the method constructs a grid over the simplex of route-choice proportions, see \ref{fig:appendix_grid_simplex}. Each point on the grid represents a global probability vector over the \(K\) available route clusters. For each grid point, agents sample valid actions according to the corresponding probabilities, subject to their individual action masks.

This sampling strategy is intended to cover a broader range of assignment patterns than purely random sampling. In particular, it includes both concentrated assignments, where most vehicles prefer one route cluster, and mixed assignments, where vehicles are distributed more evenly across alternatives.

% \begin{table}[h]
% \centering
% \caption{Assignment sampling strategies used or planned in the dataset generation pipeline.}
% \label{tab:appendix_sampling_methods}
% \begin{tabular}{lll}
% \toprule
% Method & Description & Purpose \\
% \midrule
% Greedy & Selects locally preferred candidate routes & Baseline / low-diversity assignments \\
% Random & Samples valid actions independently & Random coverage of assignment space \\
% Balanced & Encourages balanced use of available actions & Reduces route-choice imbalance \\
% Dirichlet & Samples route-choice probabilities from a Dirichlet prior & Smooth stochastic assignment variation \\
% Grid & Enumerates structured probability vectors over route clusters & Systematic assignment-space coverage \\
% \bottomrule
% \end{tabular}
% \end{table}

\section{Hyperparameters and Training Details}
\label{app:hyperparameters_training}

This appendix reports the hyperparameters used for training. The final version should include the selected hyperparameters, the search space considered, early-stopping policy, hardware setup, and total training time. Unless otherwise stated, all models are trained using mini-batch gradient descent with Adam optimisation.

We set the learning rate to 0.001, the batch size to 128, and used an LSTM as the sequence model for processing assignment matrices. For the fusion module, we used the attention-based fusion mechanism. The model was trained with the MAE loss, which directly penalises absolute deviations between predicted and simulated traffic flows.

\section{Computational Resources}
\label{app:computational_resources}

This appendix reports the computational resources required for data generation and model training. Table~\ref{tab:hardware} reports specification of hardware used in the experiments. Data generation is dominated by SUMO simulations, which are CPU-intensive and can be parallelised across independent assignment samples. Model training is GPU-accelerated and benefits from batched graph operations over the spatial cells. All training experiments were run on a workstation equipped with the GPU specified below, 16 CPU cores, and 32 GB of RAM. The average runtime was approximately 326 seconds per epoch for models with assignments branch. These details specify the compute worker type, available memory, and execution time required to reproduce the experiments.

\begin{table}[ht]
\centering
\caption{Summary of computational environment used for experiments.}
\begin{tabular}{@{}ll@{}}
\toprule
\textbf{Component}     & \textbf{Specification} \\
\midrule
CPU                    & AMD Ryzen Threadripper PRO 7995WX 96-Cores \\
GPU                    & NVIDIA GeForce RTX 4080 \\
RAM                    & 32 GB allocated per job \\
Operating system       & Ubuntu 24.04.4 LTS (GNU/Linux 6.8.0-106-generic x86\_64) \\
SUMO version           & 1.26.0 \\
\bottomrule
\end{tabular}
\label{tab:hardware}
\end{table}

\end{document}